\documentclass[conference]{IEEEtran}
\IEEEoverridecommandlockouts
\usepackage{cite}
\usepackage{amsmath,amssymb,amsfonts}
\usepackage{algorithmic}
\usepackage{graphicx}
\usepackage{textcomp}
\usepackage{xcolor}
\usepackage{booktabs}
\usepackage{multirow}
\usepackage{url}
\usepackage{subfigure}
\usepackage{bm}

\def\BibTeX{{\rm B\kern-.05em{\sc i\kern-.025em b}\kern-.08em
    T\kern-.1667em\lower.7ex\hbox{E}\kern-.125emX}}
\begin{document}

\title{ Large-scale Deterministic Transmission among IEEE 802.1Qbv Time-Sensitive Networks\\
}

\author{\IEEEauthorblockN{Weiqian Tan\IEEEauthorrefmark{1},
Binwei Wu\IEEEauthorrefmark{2}, Shuo Wang\IEEEauthorrefmark{2}\IEEEauthorrefmark{3} and Tao Huang\IEEEauthorrefmark{2}\IEEEauthorrefmark{3}}
\IEEEauthorblockA{\IEEEauthorrefmark{1}School of Cyber Science and Engineering, Southeast University, Nanjing, China\\
\IEEEauthorrefmark{2}Purple Mountain Laboratories, Nanjing, China\\
\IEEEauthorrefmark{3}State Key Laboratory of Networking and Switching Technology, BUPT, Beijing, China\\
Email: tanweiqian@seu.edu.cn, wubinwei@pml.com.cn}}

%

\maketitle

\begin{abstract}
IEEE 802.1Qbv (TAS) is the most widely used technique in Time-Sensitive Networking (TSN) which aims to provide bounded transmission delays and ultra-low jitters in industrial local area networks.
With the development of emerging technologies (e.g., cloud computing), many wide-range time-sensitive network services emerge, such as factory automation, connected vehicles, and smart grids. Nevertheless, TAS is a Layer 2 technique for local networks, and cannot provide large-scale deterministic transmission. To tackle this problem, this paper proposes a hierarchical network containing access networks and a core network. Access networks perform TAS to aggregate time-sensitive traffic. In the core network, we exploit DIP (a well-known deterministic networking mechanism for backbone networks) to achieve long-distance deterministic transmission.
Due to the differences between TAS and DIP, we design cross-domain transmission mechanisms at the edge of access networks and the core network to achieve seamless deterministic transmission. We also formulate the end-to-end scheduling to maximize the amount of accepted time-sensitive traffic. Experimental simulations show that the proposed network can achieve end-to-end deterministic transmission even in high-loaded scenarios.
\end{abstract}


\section{Introduction}

A collection of IEEE 802.1 Ethernet standards, known as Time-Sensitive Networking (TSN) \cite{DBLP:journals/corr/abs-1905-08478}, is widely applied to provide deterministic transmission (i.e., zero packet loss, bounded end-to-end delays, and ultra-low jitters) in industrial local area networks \cite{8870295}. With the development of emerging technologies, such as cloud computing, fog/edge computing, and Virtual Reality (VR)/Augmented Reality (AR), the Internet is needed to become a more reliable infrastructure. Many wide-range time-sensitive network services are necessary in emerging areas, like factory automation, connected vehicles, and smart grids \cite{grossman2019deterministic}. Due to the strict requirements for time synchronization and short link distances, TSN is not suited for providing wide-range deterministic transmission, and further supporting wide-range time-sensitive network services. Thus, we need to design a new network containing the existing TSN networks to achieve long-distance end-to-end deterministic transmission.

Within TSN, the IEEE 802.1Qbv (Time-Aware Shaper, TAS) \cite{8613095} is applied broadly. It leverages timed gates that open/close according to a prescribed schedule, allowing frames full access to the egress link with zero interference from other queues. In \cite{craciunas2016scheduling}, a TAS-capable switch with scheduled egress queues is designed to provide zero jitter and deterministic end-to-end latencies for software tasks. For achieving large-scale deterministic transmission, the IETF Deterministic Networking (DetNet) group promotes the standardization of deterministic techniques in Layer 3. Several candidate techniques are proposed, such as Cycle Specified Queuing and Forwarding (CSQF) \cite{chen-detnet-sr-based-bounded-latency-01} and Deterministic IP (DIP) \cite{qiang-detnet-large-scale-detnet-05}.
Our prior work \cite{wang2021large} deploys DIP in real large-scale networks and proves the effectiveness in large-scale deterministic transmission. In research efforts, the Paternoster algorithm \cite{seaman2019paternoster} uses four queues that alternate every \textit{epoch} using only frequency synchronization. However, Paternoster still lacks the analysis of effectiveness.
iTSN \cite{lee2017new} is a methodology for interconnecting multiple TSN networks for V2X communication.
However, the communication distance of iTSN is only 1-2 km, which cannot be applied to provide deterministic transmission among TSN networks distributed across a wide area. Besides, all of the above-mentioned large-scale deterministic transmission mechanisms just focus on packet-level transmission. They cannot achieve application-level deterministic transmission.

Based on the research actuality and to tackle the problems in previous works, we propose a hierarchical network and achieve application-level scheduling on it.
    The hierarchical network contains local area access networks and a large-scale core network. Access networks perform TAS, while the core network uses DIP to provide large-scale deterministic transmission.
    In source hosts, we extend task-level scheduling in \cite{craciunas2016scheduling} and construct the relationship between application messages and packets, Besides, a traffic shaping mechanism is proposed to increase the number of scheduled applications.
    At the edge of TAS-D and DIP-D, we design cross-domain mechanisms to guarantee deterministic transmission between TAS and DIP.
    The optimization of end-to-end scheduling is formulated as a Mixed-Integer Programming (MIP) problem. The target of the optimization is to maximize the number of accepted applications.
    We design simulation experiments to prove the proposed network and transmission mechanisms are effective to achieve large-scale deterministic transmission among TAS networks, and the optimization can improve the number of scheduled applications.

\section{Background} \label{sec2}

\subsection{Time-Aware Shaper (TAS)}
\begin{figure}[]
\centering
\includegraphics[width=.50\linewidth]{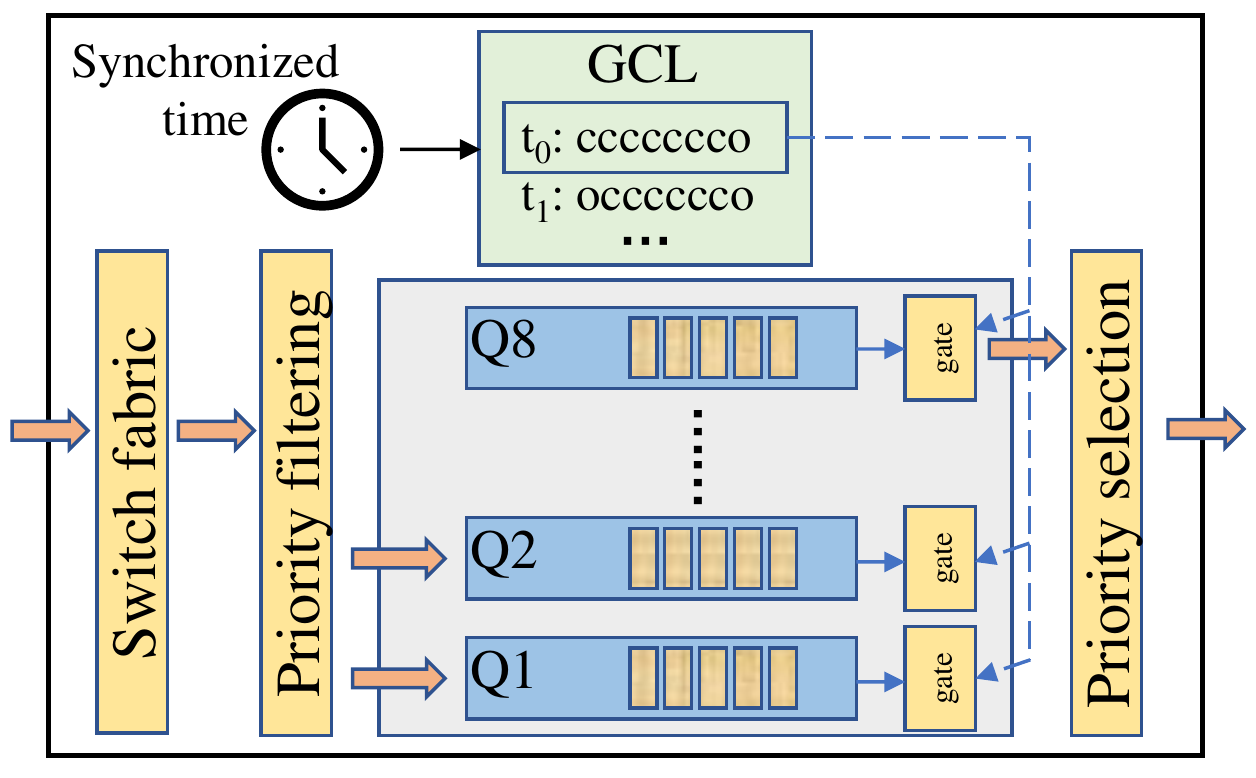}
\caption{
The internals of a TAS switch.
}
\label{fig::TASSwitch}
\end{figure}

In this section, we briefly introduce the functionality of IEEE 802.1Qbv (TAS).
Fig.~\ref{fig::TASSwitch} depicts the internals of a TAS switch.
Incoming packets pass through switching fabric and are redirected to the desired egress port. Then the priority filter assign packets to different queues (Q1 to Q8) based on the priority code point (PCP) of the IEEE 802.1Q header. Every queue at the egress port buffers packets in FIFO order, and it is associated with a timed gate. An open/close instruction of a timed gate is referred to as a Gate Control Entry (GCE). A GCE dictates which queues are allowed to access transmission medium. The entire cyclic sequence of GCEs is named Gate Control List (GCL).
In Fig.~\ref{fig::TASSwitch},
the GCE at time instant $t_0$ determine that Q8 is open for transmission and Q1-Q7 are close for receiving. The duration of one GCL cycle is defined in the \textit{cycle time} ($T_{\rm{ct}}$). The execution of all GCLs in a TSN domain is synchronized (i.e., the length and start time of $T_{\rm{ct}}$ in all TAS switches are consistent) via the Precision Time Protocol (PTP) \cite{eidson2002ieee}.

\subsection{Deterministic IP (DIP)}
DIP offers deterministic transmission by assigning packets to specific cycles in every hop. Time in network devices performing the DIP mechanism (DIP routers) is divided into cycles with length $T_{\rm{dip}}$. The cycles in a DIP-enabled network are frequency synchronized. For a pair of adjacent DIP routers $(v_1, v_2)$, assume that the packets sent in the cycle $x$ by the upstream node $v_1$ will arrive at the downstream node $v_2$ no later than the cycle $y$ (based on the propagation delay between $v_1$ and $v_2$). Then these packets will be retransmitted by $v_2$ in the cycle $(y+1)$, and the cycle mapping relationship $x \rightarrow (y+1)$ is established.

\begin{figure}[]
\centering
\includegraphics[width=.65\linewidth]{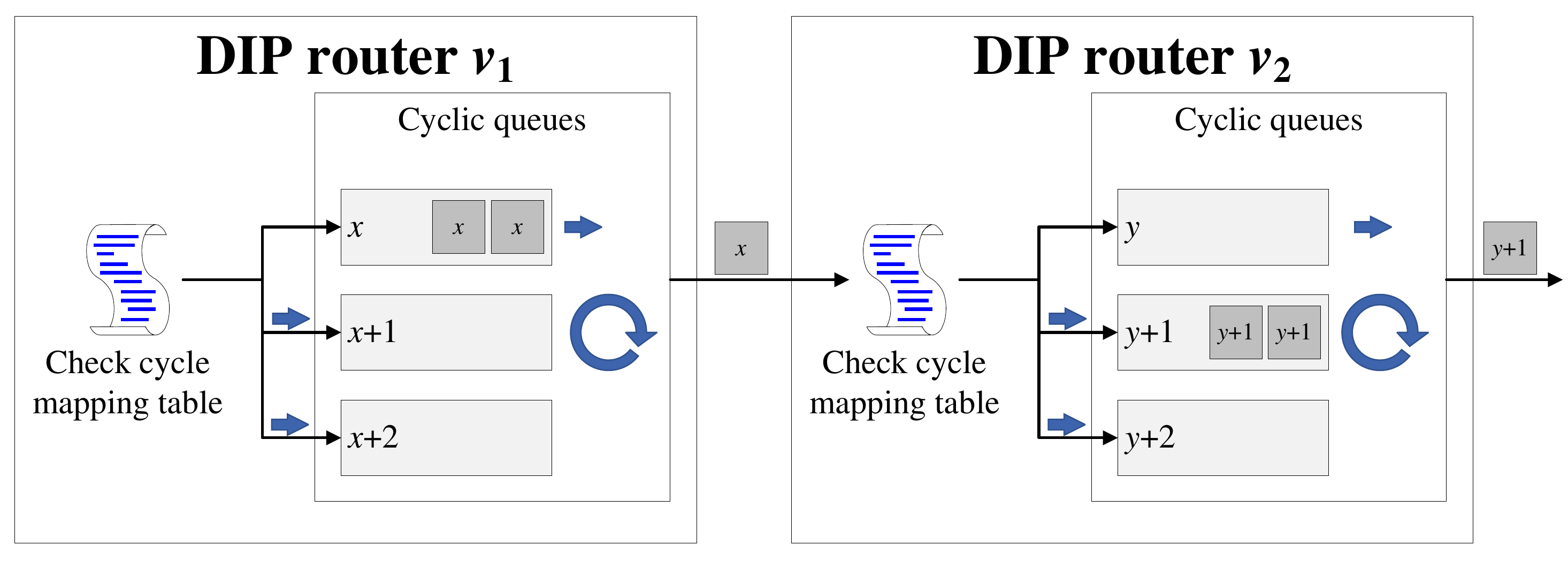}
\caption{
The transmission between a pair of DIP routers.
}
\label{fig::DIProuter}
\end{figure}

Fig.~\ref{fig::DIProuter} shows the implementation of the pair of DIP routers $(v_1, v_2)$. Each queue in a egress port is corresponding to a cycle. During a specific cycle, the queue corresponding to the cycle is open for transmission, while others are close for reception. Take the cycle mapping $x \rightarrow (y+1)$ as an example. A packet sent in the cycle $x$ by the node $v_1$ carries a identifier $x$. When the node $v_2$ finishes receiving this packet, $v_2$ will check the cycle mapping table and find $x \rightarrow (y+1)$. Then, this packet is assigned to the egress queue $(y+1)$ and replaces the identifier $x$ with $(y+1)$. When the cycle $(y+1)$ comes, the queue $(y+1)$ is open.

\section{Network design} \label{sec3}

\subsection{The overview of the hierarchical network}
\begin{figure}[]
\centering
\includegraphics[width=.65\linewidth]{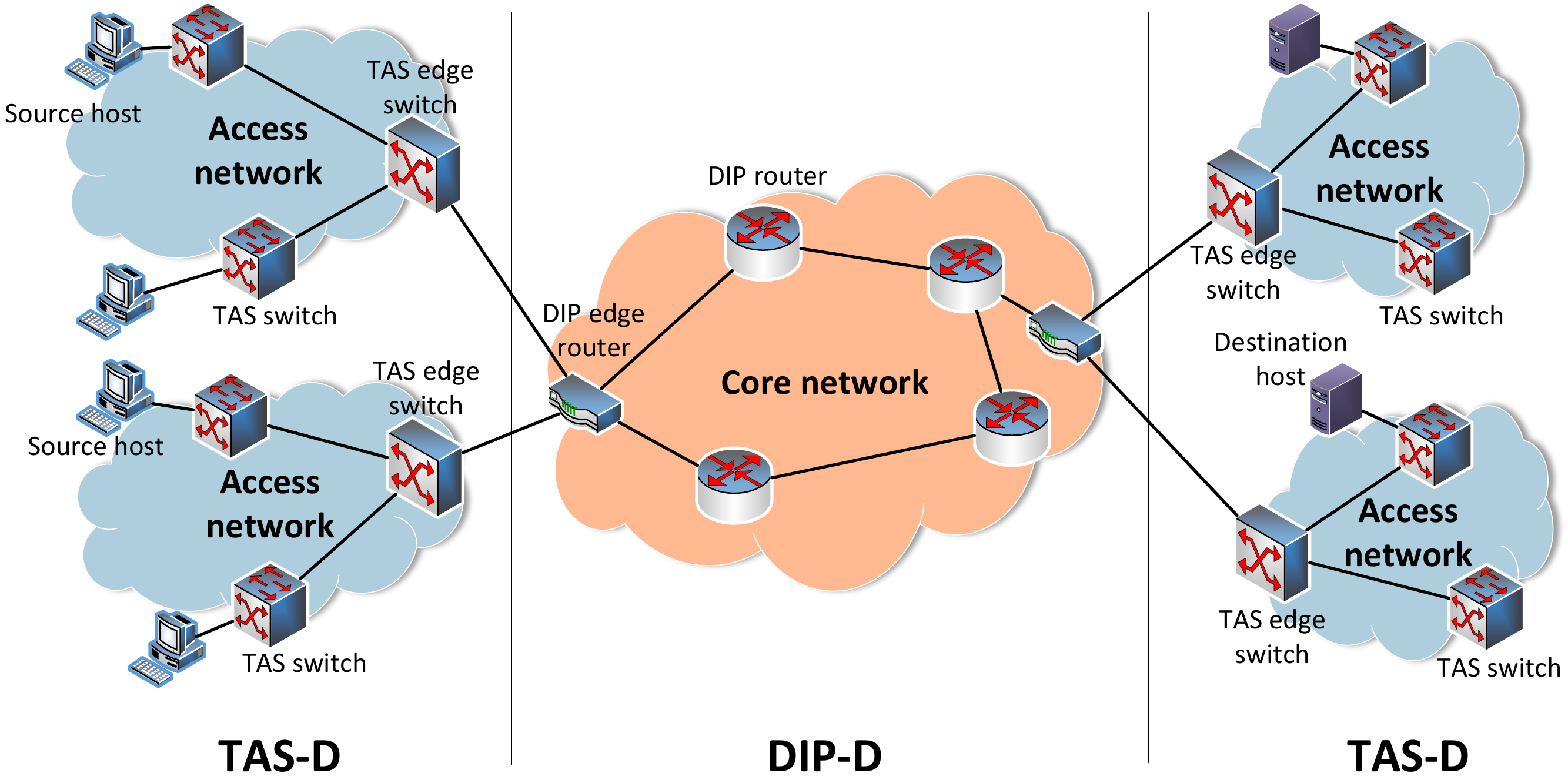}
\caption{
The hierarchical network. Access networks perform TAS and thus are logically classified into TAS-D. Similarly, the core network performing DIP is classified into DIP-D.
}
\label{fig::networkModel}
\end{figure}

In this section, a hierarchical network is presented to achieve long-distance end-to-end deterministic transmission. As shown in Fig.~\ref{fig::networkModel}, the network is divided into TAS (TAS-D) and DIP (DIP-D) domains. TAS-D leverages TAS to provide deterministic traffic aggregation in access networks, and DIP-D uses DIP to support long-distance deterministic transmission across the large area core network. TAS-D contains source/destination hosts and TAS (edge) switches, while DIP-D contains DIP (edge) routers.
We design \textit{hypercycle} in DIP-D to unify the scheduling in TAS-D and DIP-D. In TAS, the duration of a GCL cycle is named \textit{cycle time} with a length $T_{\rm{ct}}$. The resource allocation unit in TAS-D is $T_{\rm{ct}}$. The length of a hypercycle $T_{\rm{hc}}$ is equal to $T_{\rm{ct}}$. Besides, $T_{\rm{hc}}$ is a multiple of the duration of a DIP cycle $T_{\rm{dip}}$. Therefore, $T_{\rm{hc}}$ satisfies:
\begin{equation}\label{eq::hcLen}
    T_{\rm{hc}} = T_{\rm{ct}} = N_{\rm{dip}}T_{\rm{dip}} \quad (N_{\rm{dip}} \in \mathbb{Z}+)
\end{equation}


\subsection{Network and traffic model}
The network is modeled as a directed graph $\mathcal{G}=\{\mathcal{V}, \mathcal{L}\}$. $\mathcal{V}$ is the set of nodes (containing source hosts, destination hosts, TAS switches, TAS edge switches, DIP routers and DIP edge routers, i.e. $\mathcal{V} = \{\mathcal{V}_{\rm{src}}, \mathcal{V}_{\rm{dest}},  \mathcal{V}_{\rm{tas}}, \mathcal{V}_{\rm{tas}}^{\rm{edge}},  \mathcal{V}_{\rm{dip}}, \mathcal{V}_{\rm{dip}}^{\rm{edge}}\}$). $\mathcal{L}$ is the set of links.
A link $l = (v_1, v_2)$ with $l \in \mathcal{L}$, $v_1, v_2 \in \mathcal{V}$ is uniquely identified by its source $v_1$ and end $v_2$.
It is characterized by a tuple $\langle l.bw, l.d, l.q\rangle$, where $l.bw$ is the bandwidth of the link $l$, $l.d$ is the link delay, and $l.q$ represents the number of egress queues used for deterministic transmission in the source node of link $l$ (i.e., $v_1$). The link delay $l.d$ refers to the propagation delay on the medium and the processing delay in nodes. We regard $l.d$ as a constant.

In a source host $v_0 \in \mathcal{V}_{\rm{src}}$, a time-sensitive application $\tau$ emits periodic unicast messages.
An application is described by a tuple $\langle \tau.src, \tau.dest, \tau.e2e, \tau.L, \tau.T\rangle$
where $\tau.src, \tau.dest, \tau.e2e, \tau.L$, and $\tau.T$ represent the source host, the destination host, the maximum acceptable end-to-end delay, the message size, and the period, respectively.

Because the transmission pattern in every $T_{\rm{ct}}$ is the same in TAS, $T_{\rm{ct}}$ must be a multiple of the period of the messages of an application $\tau$ (i.e., $T_{\rm{ct}} = N_{\tau,m} \cdot \tau.T$, $N_{\tau,m} \in \mathbb{N}^+$). The messages within a $T_{\rm{ct}}$ are denoted by $\mathcal{M}_{\tau} = \{m_{\tau,i} | i \in [1,N_{\tau,m}]\}$.

In the transmission medium, a message may be divided into multiple packets. The number of packets belonging to the same message of $\tau$ is denoted by $N_{\tau,p}$. The set of packets belonging to the application $\tau$ is expressed as $\mathcal{P}_{\tau} = \{ p_{\tau,i,j} | i \in [1,N_{\tau,m}], j \in [1,N_{\tau,p}] \}$. The packet size of a packet $p_{\tau,i,j}$ is denoted by $p_{\tau,i,j}.L$.

\begin{figure}[]
\centering
\includegraphics[width=.65\linewidth]{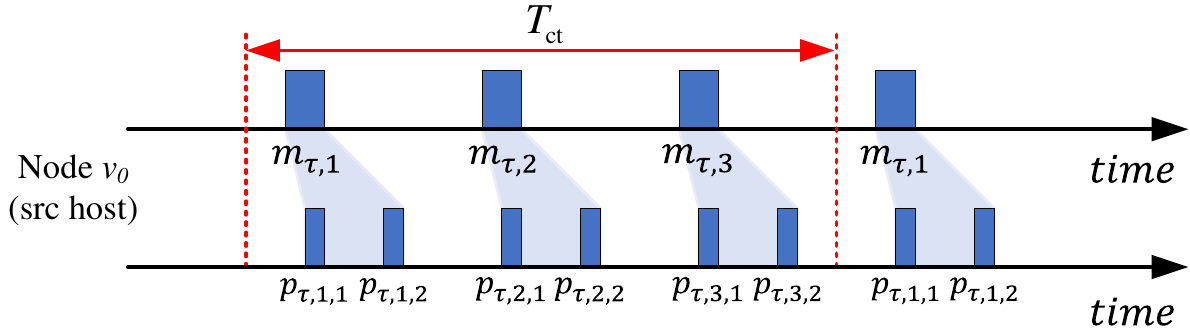}
\caption{
The relationship between messages and the corresponding packets. A cycle time $T_{\rm{ct}}$ contains 3 messages, and each message is divided into 2 packets for transmission.
}
\label{fig::msg2pkt}
\end{figure}

The relationship between messages of the application $\tau$ in the source host $v_0$ and the corresponding packets are shown in Fig.~\ref{fig::msg2pkt}. In this example, $N_{\tau,p} = 2$, and $N_{\tau,m} = 3$.




\addtolength{\topmargin}{0.01in}
\subsection{Traffic shaping in source hosts}
In source hosts, the start time of an application's transmission is random. Messages belonging to different applications may arrive at the same time. Thus, we need traffic shaping in source hosts to provide deterministic transmission.

For a message $m_{\tau,i}$, the offset between the arriving time and the start time of $T_{\rm{ct}}$ is denoted by $m_{\tau,i}.\phi$. If a packet $p_{\tau,i,j}$ is derived from the message $m_{\tau,i}$, the offset between the sending time of $p_{\tau,i,j}$ and the start time of $T_{\rm{ct}}$ is expressed as $p_{\tau,i,j}.\phi_{v_0}$, where $v_0$ is the source host. We forbid the transmission across two cycle time in TAS, so $p_{\tau,i,j}.\phi_{v_0} \in [0,T_{\rm{ct}} - \frac{p_{\tau,i,j}.L}{(v_0, v_{1}).bw}]$.

\begin{figure}[]
\centering
\includegraphics[width=.55\linewidth]{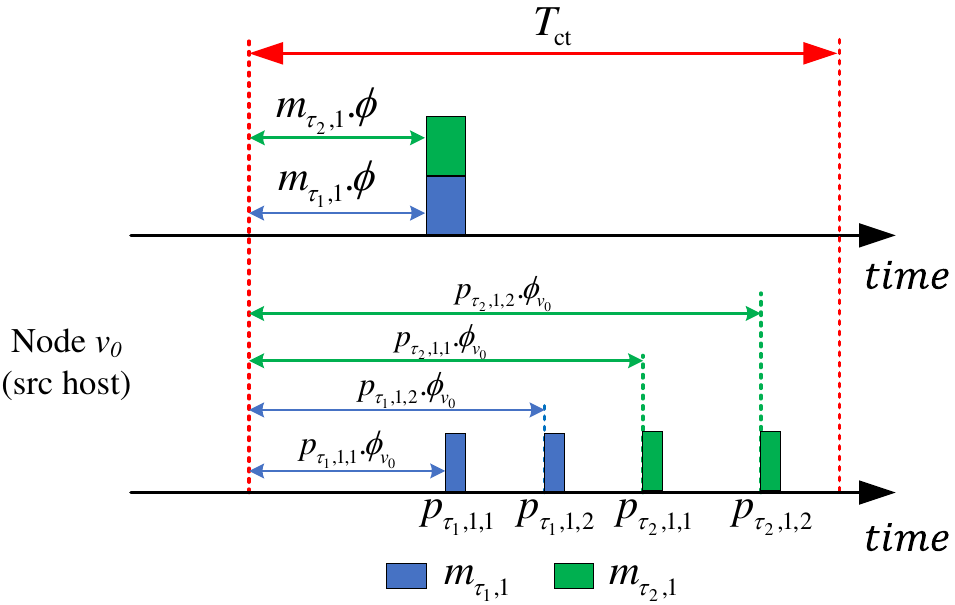}
\caption{
The traffic shaping in source hosts. Every packet is assigned a fixed time within a $T_{\rm{ct}}$ for transmission.
}
\label{fig::shapingInSrc}
\end{figure}

Fig.~\ref{fig::shapingInSrc} shows the traffic shaping of applications $\tau_1$ and $\tau_2$. The messages of $\tau_1$ and $\tau_2$ (i.e. $m_{\tau_1,1}$ and $m_{\tau_2,1}$) arrive simultaneously. We assign different offsets of packets (like $p_{\tau_1,1,1}$, $p_{\tau_2,1,1}$, etc.) to determine the transmission in the source host.

{\bf{Implementation}}: Take the example shown in Fig.~\ref{fig::shapingInSrc}. The source host $v_0$ performing TAS also has the internals shown in Fig.~\ref{fig::TASSwitch}. We assign the packet $p_{\tau_1,1,1}$ to Q8, $p_{\tau_1,1,2}$ to Q7, $p_{\tau_2,1,1}$ to Q6, $p_{\tau_2,1,2}$ to Q5. We can write the following GCEs to the GCL: ``$p_{\tau_1,1,1}.\phi_{v_0}$: ccccccco'', ``$p_{\tau_1,1,2}.\phi_{v_0}$: ccccccoc'', ``$p_{\tau_2,1,1}.\phi_{v_0}$: cccccocc'', ``$p_{\tau_2,1,2}.\phi_{v_0}$: ccccoccc''.


\subsection{Transmission from TAS-D to DIP-D} \label{sec::tas2dip}
Due to the differences between TAS and DIP, we design a cross-domain mechanism in DIP edge routers. Assume that an application $\tau$ is routed through a link $l = (v_1, v_2)$ with $v_1 \in \mathcal{V}_{\rm{tas}}^{\rm{edge}}$ and $v_2 \in \mathcal{V}_{\rm{dip}}^{\rm{edge}}$. The start time of a cycle time in $v_1$ is $t_1$, and the start time of a hypercycle in $v_2$ is $t_2$. The offset of the cycle time and the hypercycle is denoted by $l.\Delta_{\rm{h}}^{\rm{c}} = t_1 - t_2$, where $t_1 > t_2$. The packets transmitted by $v_1$ should be mapped to the transmission cycles in $v_2$. To improve the resource utilization, we introduce \textit{cycle shift} like \cite{KROLIKOWSKI202133} for traffic shaping. The next link of $l$ is denoted by $l+1$. The cycle shift of the packet $p_{\tau,i,j}$ is $r_{\tau,i,j}$,
where $r_{\tau,i,j} \in \mathbb{N} \cap [0, (l+1).q - 2]$.
The mapping relationship is denoted by $\Theta(p_{\tau,i,j})$. The definition of $\Theta(\cdot)$ is:
\begin{equation}
\begin{aligned}
\Theta(p_{\tau,i,j}) = & (\lceil\frac{p_{\tau,i,j}.\phi_{v_1} + p_{\tau,i,j}.L / l.bw + l.d + l.\Delta_{\rm{h}}^{\rm{c}}}{T_{\rm{dip}}}\rceil \\ & + r_{\tau,i,j}) \; \rm{mod}\; \textit{$N$}_{\rm{dip}}
\label{eq::T2DMapping}
\end{aligned}
\end{equation}
$p_{\tau,i,j}.\phi_{v_1}$ is the offset between the sending time of the packet $p_{\tau,i,j}$ and the start time of $T_{\rm{ct}}$ in node $v_1$. Equation \eqref{eq::T2DMapping} means the packet $p_{\tau,i,j}$ transmitted by $v_1$ will be retransmitted in cycle $\Theta(p_{\tau,i,j})$ by $v_2$. It is apparently that the range of $\Theta(p_{\tau,i,j})$ is $[0,N_{\rm{dip}} - 1]$.

\begin{figure}[]
\centering
\includegraphics[width=.65\linewidth]{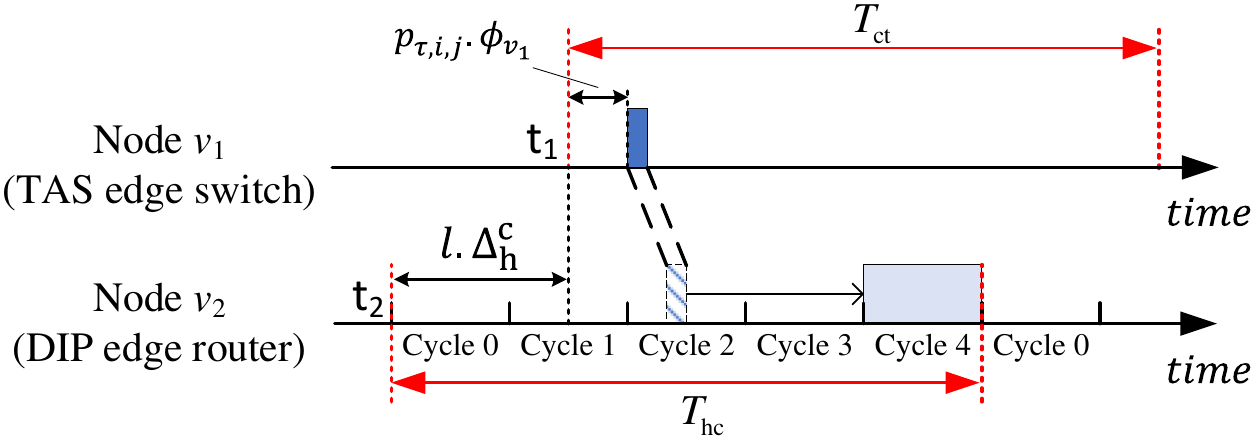}
\caption{
Transmission from TAS-D to DIP-D. A hypercycle contains 5 cycles in DIP-D. The packet $p_{\tau,i,j}$ is retransmitted in cycle 4 in Node $v_2$ (i.e., $\Theta(p_{\tau,i,j}) = 4$).
}
\label{fig::T2D}
\end{figure}

Fig.~\ref{fig::T2D} illustrates the example where $N_{\rm{dip}} = 5$, and $r_{\tau,i,j} = 1$. The packet $p_{\tau,i,j}$ will be retransmitted by $v_2$ in cycle 4 (i.e., $\Theta(p_{\tau,i,j}) = 4$).

{\bf{Implementation}}: Take the example shown in Fig.~\ref{fig::T2D}. Assume that the egress queues in $v_2$ open in a circular fashion from queue $0$ to queue $\left((l+1).q - 1\right)$. If the queue $x$ in node $v_2$ is open for transmission during cycle 2, the packet $p_{\tau,i,j}$ will be assigned to the queue $\left((x + r_{\tau,i,j} + 1)\,{\rm{mod}}\,(l+1).q\right)$.

\subsection{Transmission from DIP-D to TAS-D}
When packets come out from DIP-D, we need to form GCL in TAS edge switches. Thus, we design a cross-domain mechanism in TAS edge switches to recover the transmission in TAS-D.

A link $l = (v_1, v_2)$ with $v_1 \in \mathcal{V}_{\rm{dip}}^{\rm{edge}}$ and $v_2 \in \mathcal{V}_{\rm{tas}}^{\rm{edge}}$ is contained in the route assigned to the application $\tau$. The start time of a hypercycle in $v_1$ is $t_1$ and the start time of a cycle time in $v_2$ is $t_2$. The offset of the hypercycle and the cycle time is denoted by $l.\Delta_{\rm{c}}^{\rm{h}} = t_1 - t_2$, where $t_1 > t_2$. If a packet $p_{\tau,i,j}$ is sent in cycle $c$ by $v_1$, it will arrive at $v_2$ by $\theta_l(c)$, where $\theta_l(c) \in [0, T_{\rm{ct}})$. $\theta_l(c)$ is defined as:
\begin{equation}
\theta_l(c) = (c+1)T_{\rm{dip}} + l.d + \Delta_{\rm{c}}^{\rm{h}} - \lfloor \frac{(c+1)T_{\rm{dip}} + l.d + \Delta_{\rm{c}}^{\rm{h}}}{T_{\rm{ct}}} \rfloor T_{\rm{ct}}
\label{eq::D2T}
\end{equation}
The traffic shaping in $v_2$ is: we design an extra delay $\varphi_{\tau,i,j}$ for the packet $p_{\tau,i,j}$. The packet offset in $v_2$ is $p_{\tau,i,j}.\phi_{v_2} = \theta_l(c) + \varphi_{\tau,i,j}$. The next link of $l$ is denoted by $l+1$, and the range of $\varphi_{\tau,i,j}$ is $[0, T_{\rm{ct}})$.

\begin{figure}[]
\centering
\includegraphics[width=.65\linewidth]{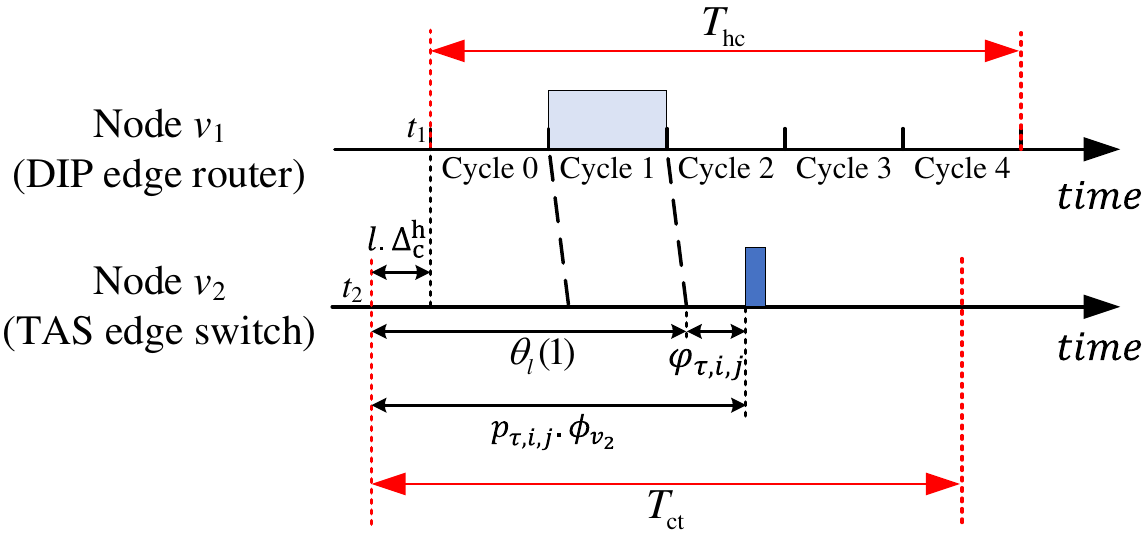}
\caption{
Transmission from DIP-D to TAS-D. The packet $p_{\tau,i,j}$ is transmitted in cycle 1 by the DIP router $v_1$, and it will be retransmitted by the TAS switch with the packet offset $p_{\tau,i,j}.\phi_{v_2} = \theta_l(1) + \varphi_{\tau,i,j}$.
}
\label{fig::D2T}
\end{figure}

Fig.~\ref{fig::D2T} shows the example where $N_{\rm{dip}} = 5$, and $c = 1$. The packet $p_{\tau,i,j}$ is all sent in cycle 1 by $v_1$, and it will arrive at $v_2$ by $\theta_l(1)$. The extra delay is $\varphi_{\tau,i,j}$, so the packet will be retransmitted with the packet offset $p_{\tau,i,j}.\phi_{v_2} = \theta_l(1) + \varphi_{\tau,i,j}$.

{\bf{Implementation}}: Take the example shown in Fig.~\ref{fig::D2T}. Since the packets' arriving order at node $v_2$ is non-deterministic, we cannot use FIFO egress queues. When a queue is open for transmission, we must ensure that the proper packet is in the first place. Thus, we leverage PIFO \cite{sivaraman2016programmable}, which allows packets to be enqueued into an arbitrary location in a queue, to guarantee the correct transmission order.

\subsection{An end-to-end example}
\begin{figure}[]
\centering
\includegraphics[width=.65\linewidth]{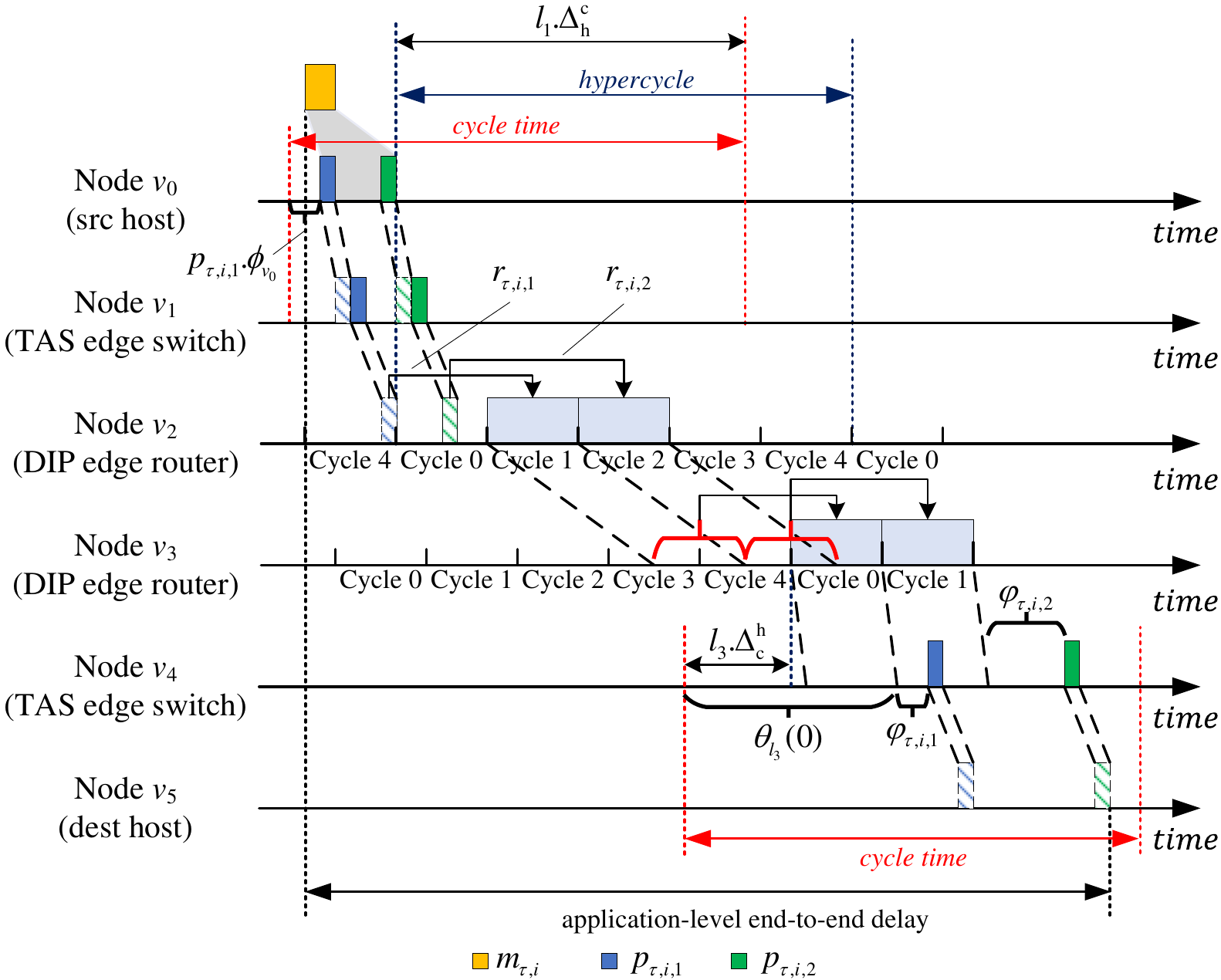}
\caption{
An end-to-end example in the hierarchical network. The route assigned to the application $\tau$ is $\bm{l}_{\tau} = (v_0, v_1, v_2, v_3, v_4, v_5)$.
}
\label{fig::e2eTrans}
\end{figure}

The end-to-end transmission of a message $m_{\tau,i}$ is illustrated in Fig.~\ref{fig::e2eTrans}. The message is divided into 2 packets ($p_{\tau,i,1}$ and $p_{\tau,i,2}$), and the period of the message is equal to $T_{\rm{ct}}$. In DIP-D, one hypercycle contains 5 dip cycles (i.e., $N_{\rm{dip}} = 5$). A route $\bm{l}_{\tau}$ is assigned to the application $\tau$. $\bm{l}_{\tau}$ is given by:
\begin{equation}
\begin{aligned}
    \bm{l}_{\tau} = & (v_0, v_1, v_2, v_3, v_4, v_5)
    \\ = & (l_0, l_1, l_2, l_3, l_4)
    \label{eq::routeExample}
\end{aligned}
\end{equation}
where $v_0 \in \mathcal{V}_{\rm{src}}$, $v_5 \in \mathcal{V}_{\rm{dest}}$, $\{v_1, v_4\} \subseteq \mathcal{V}_{\rm{tas}}^{\rm{edge}}$, and $\{v_2, v_3\} \subseteq \mathcal{V}_{\rm{dip}}^{\rm{edge}}$. Obviously, the link $l_a = (v_a, v_{a+1})$.

Take the packet $p_{\tau,i,1}$ as an example. In the source host $v_0$, the transmission offset of this packet is $p_{\tau,i,1}.\phi_{v_0}$. Thus, if $p_{\tau,i,j}$ is enqueued into the egress queue Q8, the GCL in $v_0$ must contain a GCE ``$p_{\tau,i,1}.\phi_{v_0}$: ccccccco''.
When the reception of the packet is finished in the TAS edge switch $v_1$, $p_{\tau,i,1}$ will be forwarded immediately.
After receiving the packet, the DIP edge router $v_2$ perform the cross-domain transmission mechanism. According to the cycle shift $r_{\tau,i,1} = 1$, the value of $\Theta(p_{\tau,i,1})$ is 1. Thus, $p_{\tau,i,1}$ will be transmitted in cycle 1 by $v_2$.
The transmission between two DIP edge routers $v_2$ and $v_3$ is the normal mechanism in DIP. Because the packets sent in $v_2$'s cycle 1 arrive at $v_3$ no later than cycle 4 in $v_3$, those packets are forwarded in the next cycle (i.e., cycle 0).
In the TAS edge switch $v_4$, the offset $p_{\tau,i,1}.\phi_{v_4}$ can be calculated based on $\theta_{l_3}(0)$ and $\varphi_{\tau,i,1}$. Then, $v_4$ opens the egress queue with $p_{\tau,i,1}$ according to the GCE of $p_{\tau,i,1}.\phi_{v_4}$. Finally, the destination host $v_5$ receives the packet.
When the packet $p_{\tau,i,2}$ is received by $v_5$, the transmission of the message $m_{\tau,i}$ is finished. The application-level end-to-end delay is greater than the packet-level delay of $p_{\tau,i,1}$. The transmission in every hop is accurately controlled. Thus, the hierarchical network can achieve deterministic transmission delay and zero jitters.

\section{Transmission scheduling} \label{sec4}

\subsection{Decision variables}
The decision variables of the application-level schedule contain the following:

{\bf{Admission control}:} This variable indicates whether an application $\tau$ is acceptable or not. The variable is denoted by $x_{\tau}$. If $\tau$ is accepted, $x_{\tau} = 1$. Otherwise, $x_{\tau} = 0$. Let $\mathcal{X} = \{x_{\tau} | \tau \in \Gamma\}$.

{\bf{Route selection}:} The route assigned to an accepted application $\tau$ is $\bm{l}_{\tau}$, and the definition of $\bm{l}_{\tau}$ is in \eqref{eq::route}. Let $\pounds = \{\bm{l}_{\tau} | \tau \in \Gamma\}$.

{\bf{Packet offset in Source Hosts}:} The packet offset of packet $p_{\tau,i,j}$ in the source host $v_0$ is $p_{\tau,i,j}.\phi_{v_0}$, and $p_{\tau,i,j}.\phi_{v_0} \in [0,T_{\rm{ct}} - \frac{p_{\tau,i,j}.L}{(v_0, v_{1}).bw}]$. Let $\Phi = \{p_{\tau,i,j}.\phi_{v_0} | \tau \in \Gamma, p_{\tau,i,j} \in \mathcal{P}_{\tau}, v_0 = \tau.src\}$.

{\bf{Cycle Shift}:} In the DIP edge router contained in $\bm{l_{\tau}}$, the cycle shift of packet $p_{\tau,i,j}$ is $r_{\tau,i,j}$. If the DIP edge router is the source of link $l$, $r_{\tau,i,j} \in \mathbb{N} \cap [0, l.q - 2]$. Let $\mathcal{R} = \{r_{\tau,i,j} | \tau \in \Gamma, p_{\tau,i,j} \in \mathcal{P}_{\tau}\}$.

{\bf{Extra Delay in TAS Edge Switches}:} In the edge TAS switches whose upstream node is a DIP router, the extra delay of packet $p_{\tau,i,j}$ is $\varphi_{\tau,i,j}$, and $\varphi_{\tau,i,j} \in [0, T_{\rm{ct}})$. Let $\O = \{\varphi_{\tau,i,j} | \tau \in \Gamma, p_{\tau,i,j} \in \mathcal{P}_{\tau} \}$.

\subsection{Constraints on the Transmission Mechanism}
{\bf{Conflict Constraint}:} For a link $l = (v_a, v_b)$ where $v_a \in \mathcal{V}_{\rm{tas}} \cup \mathcal{V}_{\rm{tas}}^{\rm{edge}} \cup \mathcal{V}_{\rm{src}}$, no two packets that are routed through $l$ can overlap in the time domain. For any two packets $p_{\tau_1, i_1, j_1}$ and $p_{\tau_2, i_2, j_2}$ routed through the link $l$, the constraint on the packet offsets in $v_a$ is as follows:
\begin{equation}
\begin{aligned}
& \left( p_{\tau_1,i_1,j_1}.\phi_{v_a} \geq p_{\tau_2,i_2,j_2}.\phi_{v_a} + \frac{p_{\tau_2,i_2,j_2}.L}{l.bw} \right) \vee
\\ & \left( p_{\tau_2,i_2,j_2}.\phi_{v_a} \geq p_{\tau_1,i_1,j_1}.\phi_{v_a} + \frac{p_{\tau_1,i_1,j_1}.L}{l.bw} \right)
\label{overlapConstraint}
\end{aligned}
\end{equation}

{\bf{DIP Cycle Capacity Constraint}:} For a cycle $c$ in a DIP router $v_a \in \mathcal{V}_{\rm{dip}}$ on a link $l = (v_a, v_b)$, if a set of packets $\mathcal{P}_c$ is assigned to it, the total size of $\mathcal{P}_{c}$ should not exceed the transmission capacity of $c$:
\begin{equation}
\sum_{p_{\tau,i,j} \in \mathcal{P}_{c}} p_{\tau,i,j}.L \leq T_{\rm{dip}} \cdot l.bw
\label{cycCapacityConstraint}
\end{equation}

\addtolength{\topmargin}{0.02in}
\subsection{Constraints on application-level end-to-end Delay}
We define the route assigned to the application $\tau$ as follows:
\begin{equation}
\begin{aligned}
\bm{l}_{\tau} = & (v_0, v_1, \cdots, v_k, v_{k+1}, \cdots, v_{m}, v_{m+1}, \cdots, v_{n})
\\ = & (l_0, l_1, l_2, \cdots, l_{n-1})
\end{aligned}
\label{eq::route}
\end{equation}
where $v_0 \in \mathcal{V}_{\rm{src}}$, $v_n \in \mathcal{V}_{\rm{dest}}$, $\{v_1, v_2, \cdots, v_{k-1}\}
\subseteq \mathcal{V}_{\rm{tas}}$, $\{v_k, v_{m+1}\} \subseteq \mathcal{V}_{\rm{tas}}^{\rm{edge}}$, $\{v_{m+1}, v_{m+2}, \cdots v_{n-1}\} \subseteq \mathcal{V}_{\rm{tas}}$, $\{v_{k+1}, v_m\} \subseteq \mathcal{V}_{\rm{dip}}^{\rm{edge}}$, and $\{v_{k+1}, v_{k+2}, \cdots, v_{m}\} \subseteq \mathcal{V}_{\rm{dip}}$. Obviously, the link $l_a = (v_a, v_{a+1})$.

In DIP-D, for a link $l = (v_a, v_b)$, and $v_a, v_b \in \mathcal{V}_{\rm{dip}} \cup \mathcal{V}_{\rm{dip}}^{\rm{edge}}$, the start time of a hypercycle in $v_a$ and $v_b$ is $t_a$ and $t_b$, respectively. The offset of hypercycles in $v_a$ and $v_b$ is denote by $\Delta_{\rm{h}}^{\rm{h}}(l) = t_a - t_b$ ($t_a > t_b$). A packet $p_{\tau,i,j}$ sent in cycle $c$ by $v_a$ is retransmitted by $v_b$ in cycle $\vartheta_l(c)$. The definition of $\vartheta_l(\cdot)$ is:
\begin{equation}
\vartheta_l(c) = \lceil \frac{(c+1)T_{\rm{dip}} + l.d + \Delta_{\rm{h}}^{\rm{h}}(l)}{T_{\rm{dip}}} \rceil \; \rm{mod} \; \textit{N}_{\rm{dip}}
\label{eq::DIPmapping}
\end{equation}

The transmission cycles of packet $p_{\tau,i,j}$ on $(v_{k+1}, v_{k+2}, \cdots, v_{m})$ are denoted by $\bm{c} = (c_{k+1}, c_{k+2}, \cdots, c_{m})$. Based on \eqref{eq::T2DMapping} and \eqref{eq::DIPmapping}, the cycles can be calculated recursively:
\begin{equation}
c_{a} = \left\{
\begin{array}{rcl}
\Theta(p_{\tau,i,j}) & & a = k+1\\
\vartheta_{l_{a-1}}(c_{a-1}) & & a \in [k+2,m]
\end{array} \right.
\label{eq::cycle}
\end{equation}

The packet-level end-to-end delay of $p_{\tau,i,j}$ is denoted by $\Delta_{\tau,i,j}(\bm{l}_{\tau})$. The value of $\Delta_{\tau,i,j}(\bm{l}_{\tau})$ is:
\begin{equation}
\begin{aligned}
\Delta_{\tau,i,j}(\bm{l}_{\tau}) = & p_{\tau,i,j}.\phi_{v_0} - m_{\tau,i}.\phi + \sum_{a = 0}^{k-1}\left(\frac{p_{\tau,i,j}.L}{l_a.bw} + l_a.d\right)
\\ & + \lceil \frac{p_{\tau,i,j}.\phi_{v_k} + \frac{p_{\tau,i,j}.L}{l_k.bw} + l_k.d + l_k.\Delta_{\rm{h}}^{\rm{c}}}{T_{\rm{dip}}} \rceil T_{\rm{dip}}
\\ & - p_{\tau,i,j}.\phi_{v_k} - l_k.\Delta_{\rm{h}}^{\rm{c}} + (r_{\tau,i,j} + 1) T_{\rm{dip}} + l_{k+1}.d
\\ & + \sum_{a = k+1}^{m-2} ( \lceil \frac{(c_a + 1) T_{\rm{dip}} + l_a.d + \Delta_{\rm{h}}^{\rm{h}}(l_a)}{T_{\rm{dip}}} \rceil T_{\rm{dip}}
\\ & - c_a T_{\rm{dip}} - l_a.d - \Delta_{\rm{h}}^{\rm{h}}(l_a) + l_{a+1}.d ) + \varphi_{\tau,i,j}
\\ & + \sum_{a = m}^{n - 1}\left(\frac{p_{\tau,i,j}.L}{l_a.bw} + l_a.d\right)
\label{eq::pktE2EDelay}
\end{aligned}
\end{equation}

The application-level end-to-end delay of the application $\tau$ should be less than $\tau.e2e$, the constraint is as follows:
\begin{equation}
\max_{i \in [1,N_{\tau,m}], j \in [1, N_{\tau,p}]} \Delta_{\tau,i,j}(\bm{l}_{\tau})
\leq \tau.e2e
\label{e2eDelayConstraint}
\end{equation}

\subsection{Objective Function}
The target of the scheduling is to accept as many applications as possible. In the five decision variables, the admission control and route selection are discrete variables, while the remaining three variables are continuous. Thus, the problem can be formulated as a Mixed-Integer Programming (MIP):
\begin{subequations}\label{P1}
\begin{alignat}{2}
& \max\limits_{\mathcal{X}, {\pounds}, {\Phi}, {\mathcal{R}}, {\O}} \sum_{\tau \in \Gamma} x_{\tau} &  \\
& s.t. \quad x_{\tau} \in \{0,1\}&  \\
& \quad\quad\quad \eqref{overlapConstraint},
\eqref{cycCapacityConstraint},\eqref{e2eDelayConstraint}&
\end{alignat}
\end{subequations}

\section{Simulation} \label{sec5}
In the simulation, we construct the core network in the proposed hierarchical network based on a real-world network Atlanta \cite{DBLP:journals/networks/OrlowskiWPT10}. The network is established in OMNet++, and it contains 15 DIP (edge) routers. We set 10 access networks in the simulation. Every access network contains a TAS edge switch connected to a DIP edge router, and a host is connected to the switch. In the core network, the link delay is 150 $\mu$s, and the link capacity is 10 Gbps. The link in access networks has a propagation delay 1.5 $\mu$s, and the capacity 1 Gbps.
The period of messages in the network is 1 ms or 2 ms, randomly. Besides, the message size is randomly one or two times the MTU. The duration of a DIP cycle is $T_{\rm{dip}} = 10 \, \mu \rm{s}$. Thus, the duration of a cycle time and a hypercycle is $T_{\rm{ct}} = T_{\rm{hc}} = 2 \, \rm{ms}$. In the simulation, we leverage genetic algorithm tool box embedded in Matlab to solve the problem \eqref{P1}.

\begin{figure}[]
\centering
\includegraphics[width=.65\linewidth]{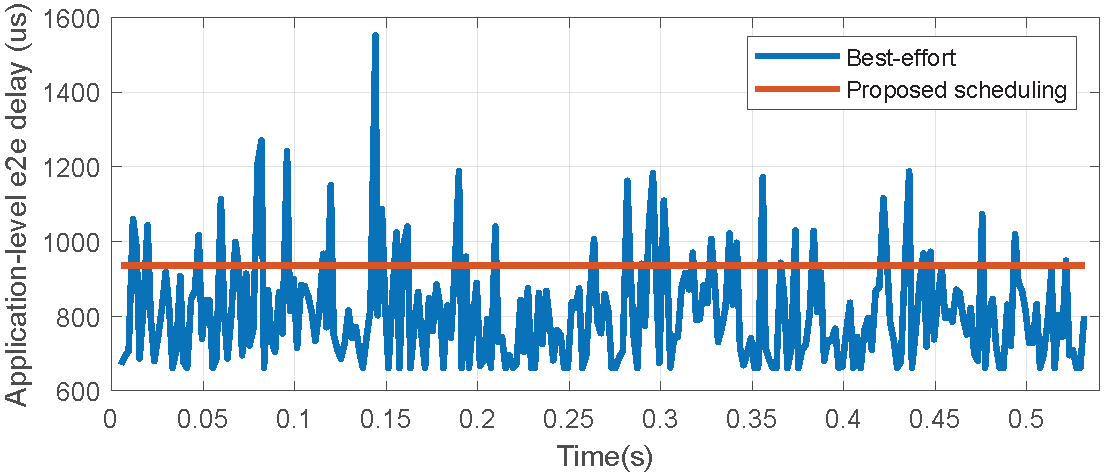}
\caption{
The application-level end-to-end delays in a medium-loaded scenario with network utilization $59 \%$.
}
\label{fig::e2eDelay}
\end{figure}

Fig.~\ref{fig::e2eDelay} shows the application-level end-to-end delays of a time-sensitive application. We rewrite the components in OMNet++ to emit interference flows for creating various congestion levels. In Fig.~\ref{fig::e2eDelay}, the network utilization is $59\%$. The end-to-end delays in best-effort transmission vary from 662 $\mu$s to 1151 $\mu$s, while the proposed scheduling can achieve deterministic end-to-end delay 953 $\mu$s with zero jitters. Moreover, the delay of proposed scheduling is larger than the minimum delay of best-effort transmission. Obviously, the control of transmission in every hop may lead to extra delays. Thus, the deterministic end-to-end delay is greater. However, the proposed scheduling can achieve zero jitters, and the delay is significantly less than the maximum delay of best-effort transmission.

\begin{figure}[]
\centering
\includegraphics[width=.45\linewidth]{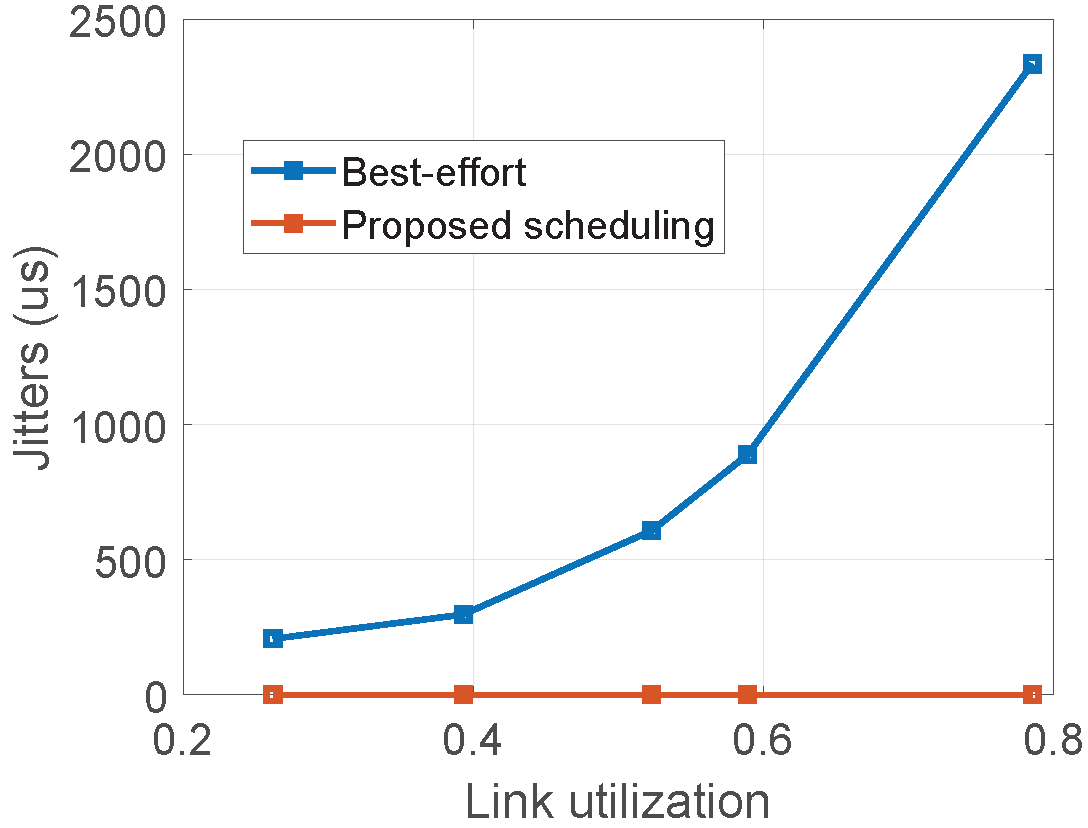}
\caption{
The transmission jitters in different congestion levels. Jitters in best-effort transmission grow significantly with the increase of network utilization. The proposed scheduling can achieve zero jitters.
}
\label{fig::jitters}
\end{figure}

Fig.~\ref{fig::jitters} depicts the transmission jitters with the increase of network utilization. We gradually increase the number of interference flows to create different congestion levels. The proposed scheduling can remain zero jitters, while the jitters of best-effort transmission grow exponentially. Due to the isolation of transmission between time-sensitive messages and interference flows, the time-sensitive messages will not interleave with interference flows. In best-effort transmission, the interleaving results in non-deterministic delays in egress queues, which create jitters.

\begin{figure}[]
\centering
\includegraphics[width=.45\linewidth]{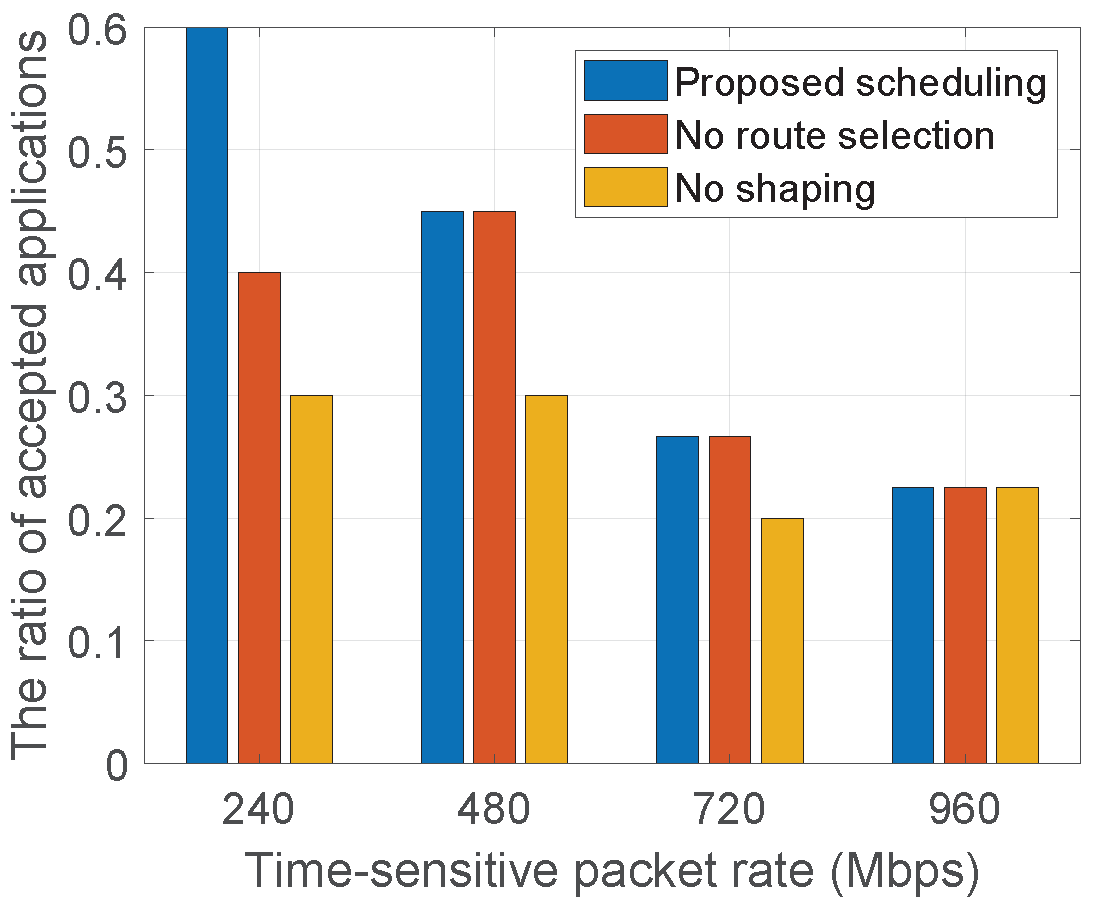}
\caption{
The ratio of accepted applications in different packet rates using different transmission mechanisms. The proposed scheduling can accept more applications for deterministic transmission.
}
\label{fig::acceptedRatio}
\end{figure}

Fig.~\ref{fig::acceptedRatio} illustrates the ratio of accepted applications using different transmission mechanisms. ``No route selection'' means that all applications perform ``shortest route first''. Moreover, ``No shaping'' implies that messages are forwarded immediately when they are created in source hosts, and the cycle shifts in DIP edge routers are always 0. The link capacity in TAS-D is 1 Gbps, and we gradually increase the packet rates of time-sensitive packets to create different levels of congestion. Compared with ``no shaping'', the proposed scheduling can accept more applications in light-loaded, medium-loaded, and high-loaded scenarios (i.e., 240 Mbps, 480 Mbps, and 720 Mbps, respectively). In extremely high-loaded scenarios (with the packet rate of 960 Mbps), the ratios of accepted applications in proposed scheduling and ``no shaping'' are the same, because there is little extra transmission resource for traffic shaping. Besides, the acceptance ratios are the same between the proposed scheduling and ``no route selection''. Because packets belonging to the same access networks have to pass through the same TAS link to DIP-D, which attenuates the influence
of route selection. In conclusion, the proposed scheduling can improve the number of accepted applications.

\section{conclusion} \label{sec6}
To empower large-scale deterministic transmission among TAS networks, this paper proposes a hierarchical network containing access networks and a core network. In access networks, we exploit TAS to provide deterministic transmission during the aggregation of traffic from source hosts. In the core network, DIP is applied to achieve large-scale deterministic transmission. To achieve end-to-end scheduling, we design a traffic shaping mechanism in source hosts, and cross-domain transmission between TAS and DIP. Moreover, the scheduling is formulated as a MIP to improve the network throughput. Simulation results show that the proposed network can achieve deterministic transmission even in high-loaded scenarios.


%

\bibliographystyle{ieeetr}
\bibliography{myref}

\vspace{12pt}

\end{document}